\documentclass[apjl]{emulateapj}
\usepackage{times}
\usepackage{color}
\def \mathbi#1{\textbf{\em #1}}
\def \msun{\rm M_\odot}

\begin{document}

\shorttitle{Do jets move?}
\title{Do jets precess... or even move at all?}
\shortauthors{\sc{Nixon \& King}} 
\author{Chris~Nixon\altaffilmark{1,2,3} \& Andrew~King\altaffilmark{2}} 
\altaffiltext{1}{JILA, University of Colorado \& NIST, Boulder CO 80309-0440, USA}
\altaffiltext{2}{Department of Physics and Astronomy, University of Leicester, University Road, LE1 7RH Leicester, UK}
\altaffiltext{3}{Einstein Fellow}
\email{chris.nixon@jila.colorado.edu}

\begin{abstract}
Observations of accreting black holes often provoke suggestions that their jets precess. The precession is usually supposed to result from a combination of the Lense--Thirring effect and accretion disc viscosity. We show that this is unlikely for any type of black hole system,
as the disc generally has too little angular momentum compared with a spinning hole to cause any significant movement of the jet direction across the sky on short timescales. Uncorrelated accretion events, as in the chaotic accretion picture of active galactic nuclei, change AGN jet directions only on timescales $\gtrsim 10^7$~yr. In this picture AGN jet directions are stable on shorter timescales, but uncorrelated with any structure of the host galaxy, as observed.
We argue that observations of black--hole jets precessing on timescales short compared to the accretion time would be a strong indication that the accretion disc, and not the standard Blandford--Znajek mechanism, is responsible for driving the jet. This would be particularly convincing in a tidal disruption event. We suggest that additional disc physics is needed to explain any jet precession on timescales short compared with the accretion time. Possibilities include the radiation warping instability, or disc tearing. 
\end{abstract}


\section{Introduction}
\label{intro}

Jets appear in all accreting systems, from protostars \citep[e.g.][]{Daviesetal1994} to AGN \citep[e.g.][]{NW1999,Kinneyetal2000}. In all cases the terminal speed of the jet is $\gtrsim$ the escape speed from the surface of the accreting object. Studies of protostellar jets usually assume that the ultimate power source is the accretion energy of the gas disc forming the star, mediated by strong magnetic fields \citep[e.g.][and references therein]{Priceetal2012}. To tap the maximum accretion energy, a jet produced in this way must come from the innermost part of the disc near the stellar surface, and so naturally gives a terminal velocity of order the escape speed. For black holes there is debate as to whether the jet driver is again the disc accretion energy (e.g. \citealt{BP1982}; \citealt{Livioetal1999}) or instead the black hole spin \citep{BZ1977}.

Observations of jets from AGN often encourage suggestions that the jets precess \citep[e.g.][]{Falcetaetal2010,Kharbetal2010,Gongetal2011,Martetal2011}. For the two suggested types of black hole jet--driving, this requires precession either of the disc plane close to the central accretor (where the jet is launched), or instead, of the black--hole spin axis. In this Letter we consider these processes, and show that precessing jets are not easy to obtain via any of the mechanisms usually invoked. The reasons are simply: (1) the angular momentum of any single realistic accretion event is always smaller than the angular momentum of the hole; and (2) the inner disc settles rapidly into a steady shape. This is aligned to the spin if $\alpha > H/R$, and a steady warp if $\alpha < H/R$. Here $\alpha$ is the Shakura--Sunyaev viscosity parameter and $H/R$ is the disc angular semithickness, and the two cases correspond to diffusive and wavelike warp propagation respectively.

\section{Lense--Thirring effect in discs with $\alpha > H/R$}
\label{LTdiff}
We briefly describe the evolution of a misaligned disc around a spinning black hole in the regime where warps propagate diffusively -- i.e. $\alpha > H/R$ \citep{PP1983}. We discuss the wavelike case ($\alpha < H/R$) in Section~\ref{wavelike}.

The diffusive case is considered at length in the literature
(e.g. \citealt{BP1975}; \citealt{Pringle1992}; \citealt{SF1996};
\citealt{LP2006} and \citealt{NK2012}). The Lense--Thirring effect of a
spinning black hole makes tilted disc orbits precess around its angular
momentum vector at a frequency $\Omega_{\rm LT} = a(R/R_{\rm g})^{-3}\Omega_{\rm
  K}\left(R_{\rm g}\right)$ (where $a$ is the Kerr spin parameter, $R_{\rm g} =
GM/c^2$ is the black hole's gravitational radius, and $\Omega_{\rm
  K}\left(R_{\rm g}\right)$ is the Kepler frequency at disc radius $R_{\rm
  g}$) which decreases strongly with radius \citep{Thirring1918,LT1918}. This
differential precession is communicated through the disc by its viscosity,
which acts to co-- or counter--align the disc with the plane of the hole. The
inner parts of the disc quickly settle in the equatorial plane of the black
hole and the outer parts remain misaligned, with the two parts joined by a
warped region. This is the Bardeen--Petterson effect \citep{BP1975} (but note
that the equations of that paper do not conserve angular momentum; see
\citealt{PP1983}).  If an external torque (e.g. from a misaligned binary
companion) maintains the tilt at the outer edge of the disc the warp can
remain stationary, but otherwise the warp propagates outwards until the entire
disc lies in the equatorial plane. The hole--disc system thus ends up aligned
(or counter--aligned) along its original total angular momentum (the vector
sum of the original spin and disc angular momenta; \citealt{Kingetal2005}). We
note that so far all calculations of the Bardeen--Petterson effect have used
\cite{SS1973} $\alpha$ discs; a demonstration of the effect for discs
explicitly driven by the magnetorotational instability (MRI) has not yet been
attempted.

The Bardeen--Petterson evolution assumes that the disc viscosity is strong enough to communicate the differential precession efficiently through the disc. Recently \cite{NK2012} and \cite{Nixonetal2012b} have shown that for realistic parameters this often does not hold. Instead the disc is torn into many distinct planes which precess almost independently of each other \citep{Nixonetal2012b}. If the disc inclination to the black hole spin is high enough this generates significantly counter-rotating disc orbits and these lead to rapid accretion \citep[cf.][]{Nixonetal2012a}. These results markedly alter the picture of how black holes accrete, and may allow for strong precession of the inner disc plane. We return to this possibility in Section~\ref{discussion}, but for the moment consider the usual Bardeen-Petterson evolution.

To discuss possible jet precessions we let $\mathbi{J}_{\rm d}$, $\mathbi{J}_{\rm h}$ and $\mathbi{J}_{\rm t} = \mathbi{J}_{\rm d} + \mathbi{J}_{\rm h}$ be the disc, hole and total angular momentum vectors respectively, with magnitudes $J_{\rm d}$, $J_{\rm h}$ and $J_{\rm t}$. During the alignment process $\mathbi{J}_{\rm h}$ precesses around $\mathbi{J}_{\rm t}$ with an {\it initial} amplitude $\theta_{\rm i}$ defined by 
\begin{equation}
  \cos\theta_{\rm i} = \frac{\mathbi{J}_{\rm h} \cdot \mathbi{J}_{\rm t}}{J_{\rm h} J_{\rm t}}
\end{equation}
This angle is small (i.e. $\mathbi{J}_{\rm t}$ and $\mathbi{J}_{\rm h}$ are in a similar direction) either when the disc is oriented in a similar direction to the hole, or when $J_{\rm d} \ll J_{\rm h}$ (and so $\mathbi{J}_{\rm h} \simeq \mathbi{J}_{\rm t}$).

It is clear that if $J_{\rm d} \ll J_{\rm h}$ alignment cannot move the hole
spin vector very far. The inner disc must quickly become anchored to the spin
plane of the hole \citep[e.g.][]{Kingetal2005}, so alignment cannot move the
inner disc very far either. So if $J_{\rm d} \ll J_{\rm h}$ the
Lense--Thirring effect cannot drive a precessing jet.

Thus if
we have the usual Bardeen--Petterson evolution, precessions are confined at best to cases where $J_{\rm d} \gtrsim J_{\rm h}$. However this still does not generate repeated jet precession. The {\it initial} amplitude of the precession can be large, since $J_{\rm t} \gg J_{\rm h}$. But the alignment and precession timescales for the disc are similar \citep{SF1996}: after only one precession time the hole is significantly aligned with the disc. This is shown explicitly
in \citet{LP2006}, who get at most a single precession of the jet (see their Figs~6 \& 11) with significant amplitude. 

We conclude that in a tilted disc propagating warps in the diffusive regime ($\alpha > H/R)$, the Lense--Thirring effect alone cannot drive repeated jet precession, unless the disc is torn into many distinct planes \citep{Nixonetal2012b}.

\subsection{Do jets move?}
\label{angmom}
We have argued above that sustained Lense--Thirring precessions are inhibited by the dynamics of the disc--hole system. We now ask how much angular momentum can be transferred from an accretion event on to a black hole. In particular, can this change its direction significantly? We derive a simple expression for $J_{\rm d} / J_{\rm h}$ and use it to consider realistic parameters for various astrophysical systems.

The disc angular momentum is
\begin{equation}
  J_{\rm d} \sim M_{\rm d}(GMR_{\rm d})^{1/2} = M_{\rm d} R_{\rm d} V_{\rm K}\left(R_{\rm d}\right)
\label{Jd}
\end{equation}
where $M_{\rm d}$ is the disc mass, $M$ is the black hole mass, $R_{\rm d}$ a characteristic radius for the disc, $V_{\rm K}$ the Keplerian velocity and $G$ is the gravitational constant. 

The spin angular momentum of a black hole with dimensionless spin parameter $a$ is \citep{KP1985}
\begin{equation}
  J_{{\rm h}} = {GM^2a\over c}
\label{Jh}
\end{equation}
where $c$ is the speed of light. Combining (\ref{Jd}) and (\ref{Jh}) gives us
\begin{equation}
  \frac{J_{\rm d}}{J_{\rm h}} = \frac{1}{a}\frac{M_{\rm d}}{M}\frac{R_{\rm d}}{R_{\rm g}}\frac{V_{\rm K}}{c}
\label{JdJh}
\end{equation}
or equivalently
\begin{equation}
  \frac{J_{\rm d}}{J_{\rm h}} = \frac{1}{a}\frac{M_{\rm d}}{M}\left(\frac{R_{\rm d}}{R_{\rm g}}\right)^{1/2},
\label{JdJh2}
\end{equation}
where $R_{\rm g} = GM/c^2\sim 10^{13}M_8$~cm is the gravitational radius (with $M_8 = M/10^8\msun$). It is clear that this ratio can take very different values for various astrophysical systems, as we now consider.

\subsubsection{Tidal Disruption Events}
In a tidal disruption event, a star on a near--parabolic orbit around a supermassive black hole fills its tidal lobe near pericenter and is torn apart. This condition implies a pericenter separation $p$ given by
\begin{equation}
p \simeq \left({M\over M_*}\right)^{1/3}R_*
\end{equation}
where the star has mass and radius $M_*, R_*$. Since $R_{\rm d} < p$ and $ M_{\rm d} < M_*$ we find
\begin{equation}
\frac{J_{\rm d}}{J_{\rm h}} < {1\over a}\left({M_*\over
    M}\right)^{5/6}\left({R_*\over R_{\rm g}}\right)^{1/2}.
\label{tide}
\end{equation}
Even in the most favorable case of a giant star ($R \sim 10^{13}$~cm), (\ref{tide}) implies a tiny ratio
\begin{equation}
\frac{J_{\rm d}}{J_{\rm h}} \la 3\times 10^{-7}M_8^{-1/2}.
\end{equation}

This makes it obvious that any observational evidence for the movement (let alone precession) of a jet in a tidal disruption event is incompatible with jet driving by the hole spin, as is central to the standard axisymmetric Blandford--Znajek mechanism. 
If instead it is assumed that the jet is driven by the inner accretion disc, this must involve physics more complex than a standard thin disc warped by the Lense--Thirring effect. Tidal disruption events may produce geometrically thick discs and therefore could propagate warps as waves (see Section 3), but this requires $\alpha$ to be unusually small \cite[cf.][]{Kingetal2007}.

\subsubsection{Black hole binaries}
This case appears slightly more promising than a tidal disruption as the black hole and the donor star have comparable masses $M_1, M_2$, with $0.1 \lesssim M_2/M_1 \lesssim 10$. However at any one instant only a small fraction of the donor star feeds the black hole and thus again we have $M_{\rm d} \ll M$. As favorable parameters we take $R_{\rm g} \approx 3\times10^6$ cm (i.e. a $10\msun$ black hole), and a large disc radius $R_{\rm d} \lesssim  10^{13}$~cm. The largest realistic disc mass is  $M_{\rm d}  \lesssim 10^{-5}\msun$ (e.g. Eq.~5.51 of \citealt{Franketal2002}, with viscosity parameter $\alpha = 0.1$ and an accretion rate $\dot M = 10^{19}~{\rm g\, s}^{-1}$ corresponding to the Eddington limit for a $10\msun$ black hole). This gives 
\begin{equation}
  \frac{J_{\rm d}}{J_{\rm h}} =
  \frac{1}{a}\frac{M_{\rm d}}{M_1}\left(\frac{R_{\rm d}}{R_{\rm g}}\right)^{1/2}
  \lesssim {10^{-3}\over a}.
\label{JdJhbhb}
\end{equation}
Thus the disc has far too little instantaneous angular momentum to cause the hole spin axis to move on a directly observable timescale. We again conclude that jet movement would imply that the jet is not driven by the black hole spin, or by the alignment of a standard thin disc warped by the Lense--Thirring effect. We note that if the disc is geometrically thick it could propagate warps as waves (see Section 3), but this requires $\alpha$ to be unusually small \cite[cf.][]{Kingetal2007}.

\subsubsection{Active Galactic Nuclei}
This case has been considered by \cite{Kingetal2008}. The main constraint on $J_{\rm d}$ is the fact that discs which are too large tend to fragment into stars under self--gravity. \cite{Kingetal2008} show that a maximal disc of this type has $J_{\rm d}/J_{\rm h} \lesssim {\rm few}\times 10^{-2} a^{-1}$ and has an instantaneous mass $\sim 10^{-3}M$, where $M$ is the SMBH mass. Thus a mass $\sim 0.01 a M$ must pass through this kind of disc, with constant orientation, to move the direction of a centrally--produced jet by $\sim 0.1$ radian. This would take at least $10^{-2}a$ Salpeter times, i.e. $\lesssim 4\times 10^5 a$~yrs, even with continuous accretion at the Eddington rate, and typically $\ga 10^7 a$~yrs if accretion is slower and slightly intermittent. If the orientation of successive accretion disc events changes randomly, as envisaged in the chaotic accretion picture of AGN \citep{KP2006,KP2007} the spin direction would perform a random walk and so deviate less from its original direction.

We again conclude that detectable jet precession is unlikely in AGN. In the chaotic accretion picture jets generally move very little for timescales $\lesssim {\rm a~few}\times 10^6$~yr. However a sequence of significant but random accretion events can move AGN jets across the sky on longer timescales ($\gtrsim 10^7$~yrs). These conclusions agree with the facts that jets with relatively stable or closely correlated directions are seen \citep[e.g.][]{Kharbetal2006}, but jet directions do not correlate at all with any features of the host galaxy \citep{Kinneyetal2000}.

\section{Lense-Thirring effect in discs with $\alpha < H/R$}
\label{wavelike}
We have argued above that Lense--Thirring precession in standard thin discs cannot be responsible for repeated precessions of jets. However it is unlikely that the innermost regions of black hole accretion discs remain thin. In this section we discuss the possibility of precession in discs with $H/R > \alpha$. We again find that repeated precession of the jet is generally unlikely, but this time not impossible. 

In Section~\ref{LTdiff} we assumed $\alpha > H/R$, so that warps propagate
diffusively. But if $\alpha < H/R$, warps instead propagate efficiently as
waves with near--sonic velocities, and are not locally damped by viscosity. It
is therefore possible that the transmission of such waves in the inner disc
region could produce a precession. However this requires quite specific
initial conditions -- i.e. that the accreting material be arranged into a
radially narrow ring, and $\alpha$ must be small. If instead the radial extent
of the disc is large, the wave induced by the Lense-Thirring effect propagates
outwards, and either never returns (on timescales of interest) or
significantly damps before returning (the wave has to reach the outer disc
edge before reflecting back inwards). \citet{Lubowetal2002} give an example
where the disc has $R_{\rm out}/R_{\rm in} = 90$ with $H/R = 0.1$ and $\alpha
= 0.05$. In this case the inner disc effectively settles into a steady shape
while the wave slowly propagates to the outer disc. As \citet{Lubowetal2002}
remark (last paragraph of their Section 4), ``the steady-state shape of the
disc close to the hole is essentially established''. The disc quickly sets up
a shape in which the internal disc torques balance the Lense--Thirring
precession torque. Thus for any precession to occur and move the jet, the
inner regions must wait for the outward propagating wave to reach a boundary
and reflect back inwards. The reflection timescale is $\sim 2R_{\rm
  out}/c_{\rm s}$ \citep[e.g.][]{NP2010} where $R_{\rm out}$ is the distance
the wave must travel and $c_{\rm s}/2$ is the wave speed \citep{PL1995}.

This reasoning is not inconsistent with the simulations of
\cite{Fragileetal2007} which suggest repeated precession of a tilted disc
around a black hole. Here the authors do not assume an $\alpha$ viscosity, but
instead simulate the MRI in an inclined thick
disc ($H/R\sim 0.2$). As is known to happen in such cases
\citep[e.g.][]{Kingetal2007}, this implies an {\it effective} viscosity
parameter ($\alpha \approx 0.01$) rather lower than implied by observations ($\alpha
\approx 0.1-0.3$). Fig. 13 of \cite{Fragileetal2007} shows the value of alpha in their computation, ranging from $\alpha\approx 0.5$ near the innermost stable circular orbit (ISCO) to $\alpha\approx 2\times 10^{-3}$ in the centre of their disc ($R=25R_{\rm g}$) to $\alpha \approx {\rm a~few}\times 10^{-4}$ in the outer parts ($R\approx 50 R_{\rm g}$). Away from the ISCO these values are far from those inferred from observations or those predicted by shearing box simulations \citep[e.g.][]{Simonetal2012}. This may well be because the simulation run time is not long enough to allow the MRI to develop fully; for example the run time is $\sim 10$ orbits at $R=25R_{\rm g}$, and only $\sim 3$ orbits at $50R_{\rm g}$. We note that the disc precession (Fig.~16 of \citealt{Fragileetal2007}) is averaged over the disc region $20R_{\rm g} < R < 50R_{\rm g}$. We also note that the timescale on which the disc is expected to reach a steady (not precessing) shape is $\sim 1/(\alpha\Omega)$ (see Eq.~4 of \citealt{Lubowetal2002}). This timescale is much longer than the runtime of the simulations showing precession. Longer runs are needed to check whether for realistic viscosities and disc sizes the repeated precession observed in \cite{Fragileetal2007} remains, rather than damping away after only a few orbits of the disc.

A thick ($H/R \gtrsim \alpha$) small ($R \ll c_{\rm s}t_{\rm damp}$) disc can in principle precess. If one can arrange a disc like this to make a sharp transition (on a scale length less than the warp wavelength) to a thin disc outside it, the wave could see this as a hard boundary and efficiently reflect back inwards. The dynamics of such a setup is largely unexplored, but since the thick region is fed by the thin region, a minimum condition is that the tilt in the thin region must be maintained. This requires extra physics, as we advocate below.

The disc geometry needed for repeated precession in the wave--like regime is feasible for a tidal disruption event, where the gas circularizes very close to the accreting black hole, and the instantaneous accretion rate can be super--Eddington. However again this is problematic: for a thick disc with $H/R \sim 0.1$ and $\alpha \sim 0.1$ the inner disc ($R < 10R_{\rm g}$) aligns after at most a few precessions (Eq.~35 of \citealt{Bateetal2000}).
 
\section{Discussion}
\label{discussion}
We have argued that the physics of standard warped discs (diffusive or wave--like) strongly suggests that the Lense--Thirring effect alone is not a promising mechanism for explaining jet precessions, except possibly in rather rare cases (see Section~\ref{wavelike}). The essential reason for this is that the accretion disc generally has total angular momentum small compared with that of the spinning black hole, strongly restricting the motion of any jet across the sky. Two alternative mechanisms, so far largely unexplored, may offer more promising ways of moving jets. 

First \citet{Pringle1996} shows that an accretion disc can be unstable to warping driven by irradiation from a central source. If there is initially a small tilt in the disc, this can grow to provide a substantial global tilt in the disc with the angle between inner and outer parts differing by up to $\Delta\theta \sim \pi$. The inner regions of the disc precess with a quasi--periodic change in inclination \citep{Pringle1997}. This mechanism uses the angular momentum induced by anisotropic scattering of the central accretion luminosity, so could potentially be more powerful than the Lense--Thirring effect.

A second possibility for large precessions of the disc plane close to the black hole is that for large disc tilts it may break into distinct planes, with only tenuous viscous communication between them. This happens when the Lense--Thirring torque is strong enough to overcome the viscous torques holding the disc together \citep{NK2012,Nixonetal2012b}. \cite{Nixonetal2012b} show that rapid precessions can occur here. We shall explore these ideas in future papers.

Finally we note that interaction of the jet with super--Eddington winds coming from the disc can also generate precession of the jet as suggested for SS433 \citep{Begelmanetal2006}. Here the jet collides with a precessing gas mass and is deflected (and slowed). The jet precession here is purely a consequence of the deflection.

\acknowledgments We thank Phil Armitage for useful discussions. Support for
this work was provided by NASA through the Einstein Fellowship Program, grant
PF2-130098. Research in theoretical astrophysics at Leicester is supported by
an STFC Rolling Grant.

\bibliographystyle{apj}

\end{document}